\documentclass[3p,,preprint,12pt]{elsarticle}
\makeatletter\if@twocolumn\PassOptionsToPackage{switch}{lineno}\else\fi\makeatother

\usepackage{soul}
\usepackage{lineno}
\usepackage{amsmath}
\usepackage{graphicx}
\usepackage{subfig}
\usepackage{array}
\usepackage{amsfonts}
\usepackage[dvipsnames]{xcolor}
\usepackage{booktabs}
\usepackage{physics}
\usepackage{cancel}
\usepackage{pifont}
\usepackage{amssymb}
\usepackage{natbib}
\usepackage{epstopdf}

\usepackage{tabulary,xcolor}
\usepackage{amsfonts,amsmath,amssymb}
\usepackage[T1]{fontenc}
\makeatletter
\let\save@ps@pprintTitle\ps@pprintTitle
\def\ps@pprintTitle{\save@ps@pprintTitle\gdef\@oddfoot{\footnotesize\itshape \null\hfill\today}}
\def\hlinewd#1{%
  \noalign{\ifnum0=`}\fi\hrule \@height #1%
  \futurelet\reserved@a\@xhline}

\AtBeginDocument{\ifNAT@numbers \biboptions{sort&compress}\fi}
\makeatother

\usepackage{ifluatex}
\ifluatex

\usepackage{fontspec}
\defaultfontfeatures{Ligatures=TeX}

\usepackage[]{unicode-math}
\unimathsetup{math-style=TeX}
\else

\usepackage[utf8]{inputenc}
\fi
\ifluatex\else\usepackage{stmaryrd}\fi

\usepackage{url,multirow,morefloats,floatflt,cancel,tfrupee}
\makeatletter
\AtBeginDocument{\@ifpackageloaded{textcomp}{}{\usepackage{textcomp}}}
\makeatother
\usepackage{colortbl}
\usepackage{xcolor}
\usepackage{pifont}
\usepackage[nointegrals]{wasysym}
\makeatletter

\def\mcWidth#1{\csname TY@F#1\endcsname+\tabcolsep}

\usepackage{lineno}

\begin{document}

\begin{frontmatter}	
\title{\textbf{
Digital image quantification of rice sheath blight: Optimized segmentation and automatic classification
}
}
\author[a1]{Da-Young Lee\ensuremath{^{*}}}
\author[a2]{Dong-Yeop Na}
\author[a3]{Yong Seok Heo}
\author[a1]{Guo-Liang Wang\ensuremath{^{*}}}

\address[a1]{Department of Plant Pathology\unskip,
    The Ohio State University\unskip, OH\unskip, Columbus\unskip, USA}

\address[a2]{School of Electrical and Computer Engineering\unskip,
    Purdue University\unskip, IN\unskip, West Lafayette\unskip, USA}

\address[a3]{Department of Electrical and Computer Engineering\unskip,
    Ajou University\unskip, Suwon \unskip, Korea}

\footnote{\ensuremath{^{*}}Corresponding authors: Guo-Liang Wang (wang.620@osu.edu) and Da-Young Lee (lee3467@purdue.edu).}

\begin{abstract}
Rapid and accurate phenotypic screening of rice germplasms is crucial in screening for sources of rice sheath blight resistance. However, visual and/or caliper-based estimations of coalescing, necrotic, ShB disease lesions are time-consuming, {\color{black}labor intensive and exposed to human rater subjectivity}. Here, we propose {\color{black} the use of RGB images and image processing techniques} to quantify ShB disease progression in terms of lesion height and diseased area{\color{black}}. {\color{black}To be specific, we developed a pixel color- and coordinate-based K-Means Clustering (PCC-KMC) algorithm utilizing Mahalanobis metric aimed at accurate segmentation of symptomatic and non-symptomatic regions within rice stem images.}
The performance of PCC-KMC was evaluated using Lin’s concordance correlation coefficient ($\rho_{c}$) by comparing its results to visual measurements of ShB lesion height (cm) and to lesion/diseased area (cm\ensuremath{^{2}}) measured using ImageJ.
Low bias {\color{black}(C\ensuremath{_{b}})} and high precision {\color{black}(r)} were observed for {\color{black} absolute lesion height (C\ensuremath{_{b}}= 0.93, r = 0.94) and absolute symptomatic area (C\ensuremath{_{b}}= 0.98, r = 0.97) studies.}
Moreover, we introduced a convolutional neural network (CNN) for the automatic annotation on clusters, termed PCC-KMC-CNN.
Our CNN was trained based on 85\%:15\% of composition for training and testing dataset from total 168 ShB-infected stem sample images, recording 92\% accuracy and 0.21 loss.
PCC-KMC-CNN {\color{black}also showed high accuracy and precision for the absolute lesion height (C\ensuremath{_{b}}= 0.86, r = 0.90) and absolute diseased area (C\ensuremath{_{b}}= 0.99, r = 0.97) studies.}
These results demonstrate that the present methodology has a great potential and promise to substitute the traditional visual-based ShB disease severity assessment.

\end{abstract}

\begin{keyword}
Rice sheath blight \sep Digital image processing\sep Plant disease quantification\sep Color- and coordinate-based K-means clustering \sep Convolutional neural network \sep Plant disease quantification
\end{keyword}

\end{frontmatter}

\section{Introduction}
Rice sheath blight (ShB) is a major fungal disease of rice which causes drastic reduction in quantity and quality of rice production \citep{Groth2008,Ogoshi1987,Tan2007}. ShB emerged as a major rice disease in recent decades due to intensified cultivation of short-statured, high-tillering and high-yielding rice varieties accompanied by increased nitrogen input \citep{BannizaHolderness2001,Slaton2003}. The causal agent of ShB, \textit{Rhizoctonia} \textit{solani} K{\"{u}}hn AG1-IA, generates water-soaked disease lesions on rice leaf sheaths at the plant-water or soil interface \citep{Ogoshi1987}. Initially being greenish-grey, the water-soaked lesions gradually matures into greyish-white ellipsoids surrounded by dark brown to black margins on the rice leaf sheaths \citep{HashibaKobayashi1996,Ogoshi1987}.
Chemical managment of ShB using fungicides have been effective in managing the disease in the field \citep{Groth2008,HashibaKobayashi1996}. However, this approach has been increasingly discouraged due to its negative consequences to the environment and humans \citep{Asins2010}.
Furthermore, the use of host resistance is known to be the most effective and economical approach \citep{Gururani2012}, preventing the persistence of the pathogen in fields which would otherwise lead to continuous ShB outbreaks \citep{Groth2008}. Nevertheless, to date, rice germplasm with complete ShB resistance has not been identified yet \citep{Jia2007,Pinson2005,Liu2009,Zeng2011}. Hence, the screening of rice germplasms in search for sources of ShB resistance is crucial\citep{Jia2013}.

Numerous approaches for ShB resistance screening for field and controlled conditions are currently available \cite{Jia2007, Jia2013, Hossain2016, sato2004mapping, zou2000mapping}. For instance, the microchamber method provides conducive environment for optimal ShB development and allows minimal environmental effects \cite{Jia2007}. Quantification of plant disease has been commonly conducted by human rater-based visual estimations of disease severity. As to phenotype for ShB resistance, visual estimation of diseased area relative to the whole plant or ruler-based measurements of the vertical distance that disease lesions have progressed (lesion length) \cite{Bock2008,  Hossain2016,  Jia2013, Singh2002}. The limitation of this traditional method of disease phenotyping method is that it is time-consuming as well as vulnerable to rater subjectivity and inaccuracy, specifically when the screening is conducted in large scales \cite{Bock2020,  Jia2007,  Jia2013,  Lee2002,  Poland2011}. Moreover, inconsistencies in evaluating and scoring in ShB disease reactions in screening experiments arise due to subjectivity of the variety of ShB disease scoring system based on visual estimations. Precise phenotypic data is crucial in screening experiments (Hossain et al., 2014), specifically for polygenic diseases resistance screenings, due to a wide varying levels of susceptibility profiles among different rice cultivar \cite{Groth2008, sha1990resistance, MarchettiBollich1991}. Consequently, utilizing repeatable and less subjective approaches will capture minor differences in terms of resistance reactions and may provide reliable and precise phenotypic data which is help find stronger and robust germplasms in resistance breeding programs.

In recent years, the use of various imaging systems and deep learning has been extensively has beenbeen shown great promise as a tool for quantifying plant diseases \citep{FERENTINOS2018,BARBEDO2019,BARBEDO2018, Mohanty2016, Barbedo2013,Mahlein2016}. These technologies along with convolutional neural networks show promise in assessing disease severity that can reduce the human errors caused by visual assessments \citep{Rousseau2013,Bock2020,TodaOkura2019,Ma2018}. A number of image processing platorms are currently available for plant disease severity assessments, including Assess \cite{Lamari2002}, SigmaPro (Systat Software, San Jose, CA) ImageJ \citep{Schneider2012}. Moreover, numerous studies on detecting rice diseases with the RGB imagery have been documented \citep{PothenPai2020,PhadikarSil2008,Yao2009}. However, none has yet been developed specifically {\color{black}targeted} for ShB phenotyping. Moreover, One crucial point which must be addressed for new plant disease assessment methods is their capability to generate accurate measurements through validation of the measurement of the proposed method \citep{madden2007study}. `Accuracy' refers the closeness of the measured value to the actual value while `precision' refers to the variability of the measurement \citep{Nutter2006,Bock2020}. Using these parameters, validation can be done by comparing the measurements of the new method to actual values or `gold-standard', such as visual-based measurements or pre-existing, image-based measurements (ImageJ) \citep{Bock2020}.

In general, plant disease detection using {\color{black}digital} image processing consists of three major steps: (i) pre-processing (background and noise removal), (ii) segmentation, and (iii) feature extraction and identification of the object of interest \citep{Gupta2017,KhiradePatil2015,Kuruvilla2016}. Among these, image segmentation is a crucial process which delineates and groups different regions of an image. One of the most commonly used image segmentation algorithms for plant disease detection is K-means clustering (KMC) \citep{MacQueen1967, Sethy2017,Al-Hiary2011,AlBashish2011}. KMC segments RGB images into `k` numbers of clusters based on the closeness of the RGB {\color{black}color space} without the consideration of spatial proximity. Consequently, this feature limits the utility of KMC to accurately detect ShB since the disease lesions gradually spreads out from the point of pathogen inoculation and do not appear in a sporadic, random pattern.
Here, we propose a segmentation approach suited for detecting and quantifying ShB in rice stems. The objective of this study was to develop a pixel color- and coordinate-based K-means clustering (PCC-KMC) algorithm which can accurately capture disease progression of ShB and quantify the extent of the disease in terms of lesion height and area.

\section{MATERIALS AND METHODS}
\subsection{Preparation of ShB-infected plant samples}
\begin{figure}[ht]
\centering
\includegraphics[width=\linewidth]{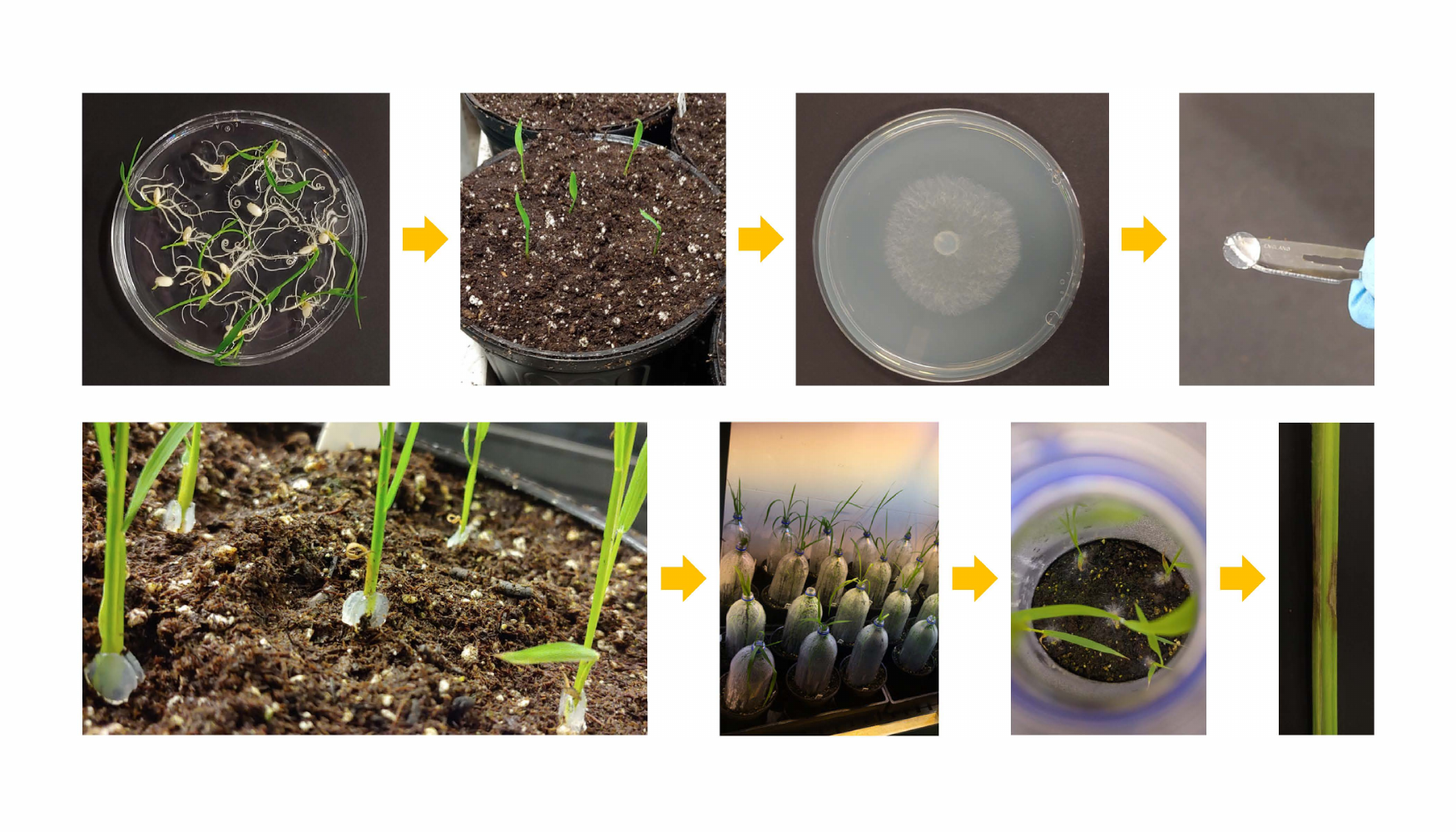}
\caption{Pictorial diagram depicting the steps in preparing rice sheath blight-infected rice stem samples for red-green-blue (RGB) imaging.}
\label{Figure1}
\end{figure}

The rice cultivars/accessions were obtained from the Genetic Stocks Oryza Collection, USDA-ARS, Dale Bumpers National Rice Research Center, Stuttgart, Arkansas \cite{Eizenga2014}. Ten rice accessions were randomly selected from the Rice Diversity Panel 1 (RDP1) (Table 1), a collection of purified, homozygous \textit{Oryza} \textit{sativa} L. accessions encompassing land races and elite rice cultivars \cite{Eizenga2014}.
\begin{table}[ht]
\centering
  \caption{List of ten RDP1 rice accessions and the number of RGB images used to for testing PCC-KMC-CNN.}
  \label{Table 1}
  \includegraphics[width=\linewidth]{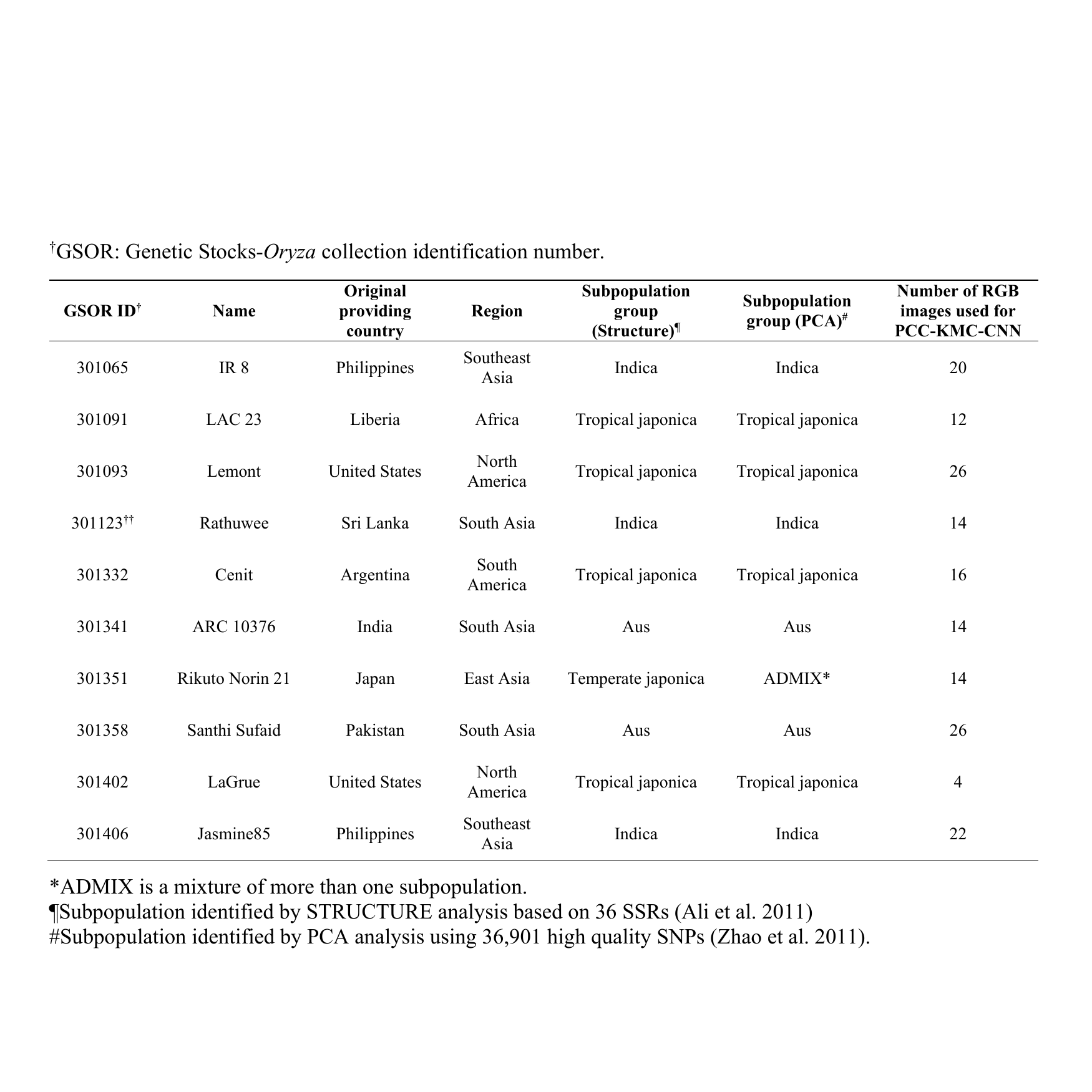}
\end{table}
De-husked rice seeds of the 10 cultivars, 301065 (IR 8), 301091 (LAC 23), 301093 (Lemont), 301123 (Rathuwee), 301332 (Cenit), 301341 (ARC 10376), 301351 (Rikuto Norin 21), 301358 (Santhi Sufaid), 301402 (LaGrue), 301406 (Jasmine 85), were surface-sterilized with 10\% Chlorox solution, germinated on 1/2 Murashige and Skoog (1/2 MS) medium then transplanted into plastic pots (15 cm in diameter) filled with ProMix BX\ensuremath{^\text{TM}} soil (Fig. 1).
The seedlings were grown in a growth chamber (26 \ensuremath{^\circ}C on day and 20 \ensuremath{^\circ}C at night, 12 hr light/dark cycle, 80\% relative humidity) until they were ready for inoculation (Fig. \ref{Figure1}).
Rice seedlings with four to five leaves were then inoculated with the ShB pathogen \textit{R}. \textit{solani} (isolate B2 provided by Dr. Jim Correll, University of Arkansas, USA).
Circular agar blocks (5 mm diameter) were excised from the border of an actively growing 3 day-old culture on potato dextrose agar (PDA).
The agar blocks were attached on both sides of the stem base of each seedling (two agar blocks per seedling).
Microchamber method \cite{Jia2007} was used  and inoculated seedlings were placed in growth chambers at 28-30\ensuremath{^\circ}C with a relative humidity of $\sim$90\% to provide optimum conditions for ShB development.
At 7 to 10 days post-inoculation, the distance between the point of pathogen inoculation (base of the rice plant) to the margin of lesion farthest from the base of the stem was measured using a caliper to reflect the disease severity \cite{HashibaKobayashi1996,  Ogoshi1987, IRRI2002, Jia2007}. After collecting visual-based measurements of ShB progression, the stem of the rice plant was excised to acquire RGB digital images.

\subsection{Acquisition of RGB digital images of ShB-infected rice stems}
All digital RGB images of ShB-infected rice stems were acquired using a flatbed scanner (HP Scanjet G4050) under controlled conditions to ensure uniformity of image acquisition.
To reduce potential background noise in the RGB image, a monochromatic, matt black paper was used as background.
Images with 600 pixels per inch (ppi) resolution were saved in PNG format.
{\color{black}
Subsequent digital image processing was performed with the use of image toolbox in MATLAB \cite{MATLAB2019}.
}

\subsection{Image analyses of ShB-infected rice stems}
\subsubsection{Pre-processing of acquired RGB images}
\begin{figure}
\centering
\includegraphics[width=\linewidth]{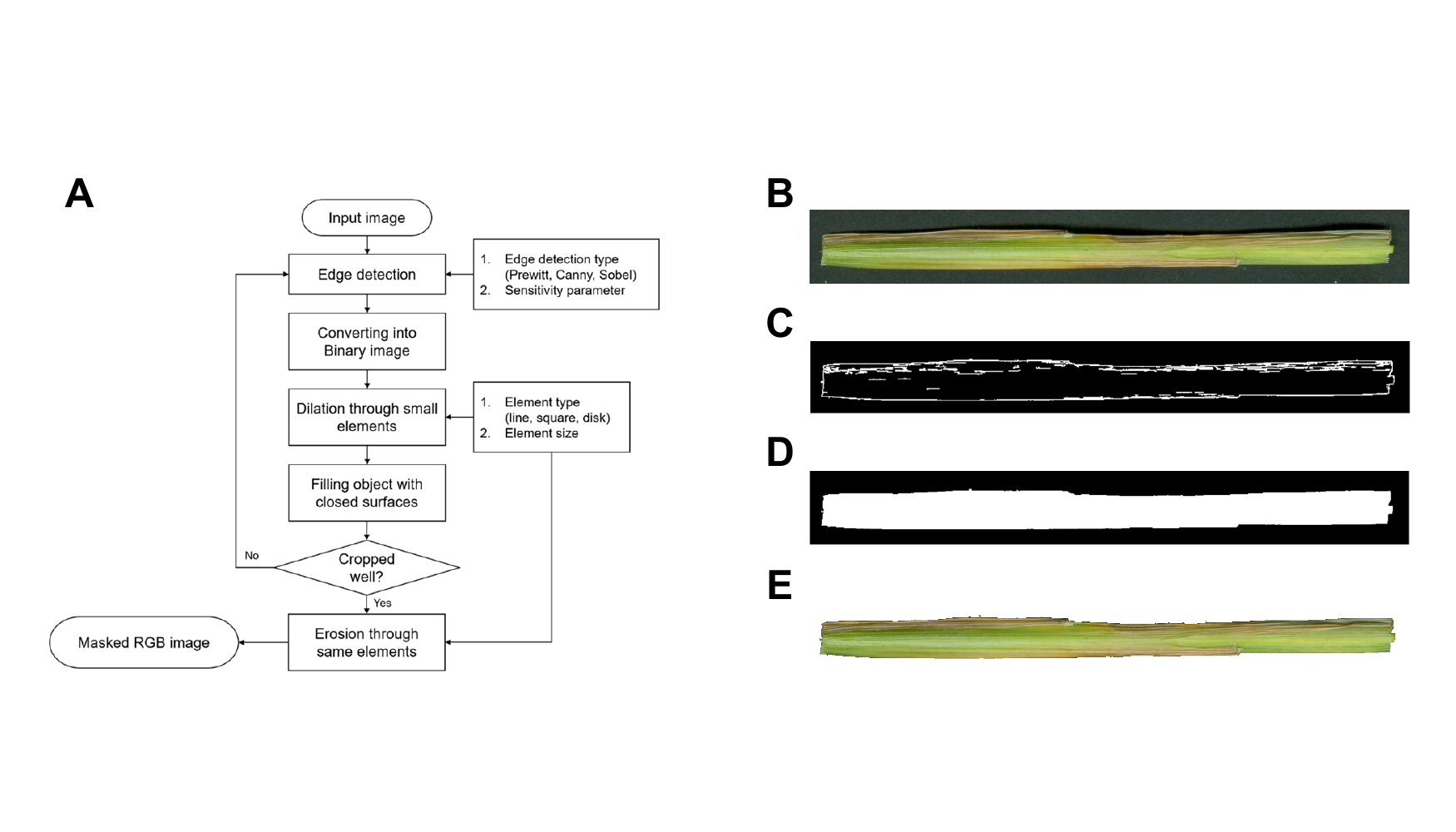}
\caption{Pre-processing steps of red-blue-green (RGB) images to isolate the plant sample of interest. {\bf A}, Schematic diagram of the background removal steps employed in this study. {\bf B}, Input RGB digital image of the stem of a sheath blight-infected rice plant (cultivar Jasmine 85). {\bf C}, Binary images after using the edge detection with dilation. {\bf D}, Binary images after filling and erosion. {\bf E}, The masked RGB digital image after {\bf C} and {\bf D} were completed.}
\label{Figure2}
\end{figure}

The acquired RGB images were all pre-processed to remove the background and to isolate the region of interest (rice stem image) (Fig. \ref{Figure2}{\bf A}).
Upon isolation of the stem image, green-dominant pixels (green, non-symptomatic tissue) were extracted through initial screening using the following {\color{black}lenient} threshold criterion:
\begin{flalign}
I_{G}(r,c) > I_{R}(r,c),\quad
I_{G}(r,c) > I_{B}(r,c),
\end{flalign}
where $I_R$, $I_G$, and $I_B$ represent red, green, and blue values of a given input RGB digital image, respectively, whereas r and c indicate row and column pixel index, respectively.
Pixels which met the above condition were labeled as 1, otherwise as 0, thereby generating two binary images/masks to decompose the input RGB into green tissue (non-symptomatic tissue) and disease lesions (symptomatic tissue).
For instance, the acquired RGB image of the stem of Jasmine 85 (Fig. 2B) was pre-processed using two key parameters associated with the performance of the background remover: (i) the edge detection type with a sensitivity parameter (Fig. 2C) and (ii) the element type and size for dilation and erosion (Fig. 2D). Consequently, successful isolation of the rice stem from the background was achieved (Fig. 2E) and the well-masked RGB image of Jasmine 85 was then subjected to color thresholding that decomposed the rice stem into regions without ShB disease symptoms (green regions) and with disease symptoms (necrotic lesions) (Fig. 3A).
\begin{figure}
\centering
\includegraphics[width=\linewidth]{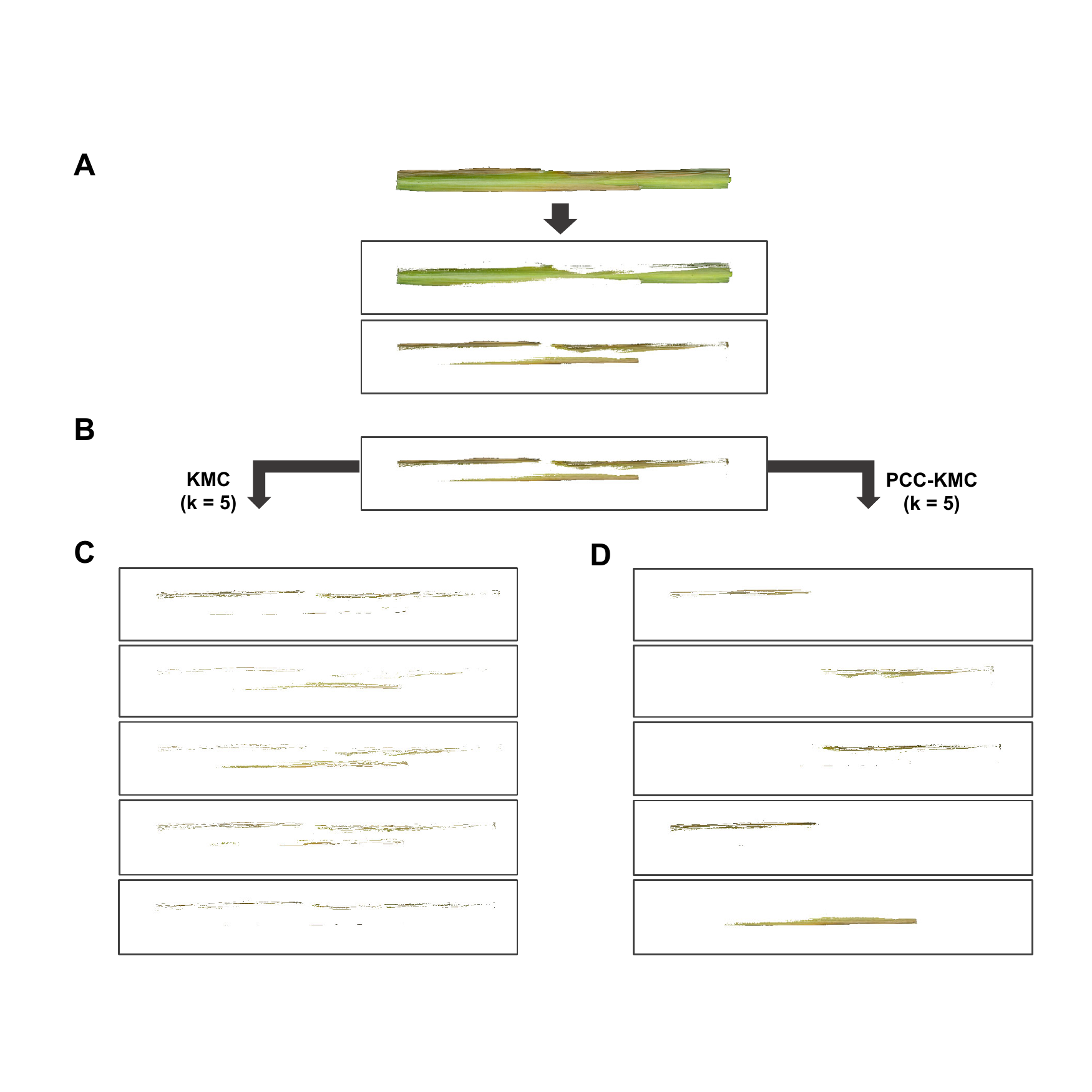}
\caption{Image segmentation of the stem of a sheath blight (ShB)-infected rice plant (cultivar Jasmine 85). {\bf A}, Segmentation of the ShB-infected Jasmine 85 sample using a simple thresholding method to partition plant regions without disease symptoms (green pixels) from those with disease symptoms. {\bf B}, Working principle for image segmentation used in KMC and PCC-KMC algorithms. Pattern of pixel clustering using {\bf C}, KMC and {\bf D}, PCC-KMC.}
\label{Figure3}
\end{figure}

\subsubsection{Segmentation of RGB images through PCC-KMC}

Since ShB gradually progresses towards the rice canopy and infect leaf blades, leaving disease lesions along its progression path (vertical and horizontal spread) \citep{HashibaKobayashi1996}, we developed an optimal segmentation method which accounts for spatial proximity as well as color similarity while clustering. We developed a pixel color- and coordinate-based K-Means clustering algorithm (PCC-KMC) operating over 5-dimensional space composed of RGB color values and row and column pixel index, i.e., red, green, blue, row, and column, wherein the measure of the distance between two data points directly affects the result of k-means clustering. Hence, instead of the naive Euclidean distance, we utlilized the Mahalanobis metric which can properly account for the scaling difference in each axis , encoded in the form of covariance matrix \cite{Melnykov2014}, in order to accurately define the closeness between a pair of data points over the 5-dimensional space. That is, for arbitrary two data points $\mathbf{x}_1=\left\{R_1,G_1,B_1,r_1,c_1\right\}^T$ and $\mathbf{x}_2=\left\{R_2,G_2,B_2,r_2,c_2\right\}^T$, one can evaluate the Mahalanobis distance between two data points as
{\color{black}
\begin{flalign}
d_{M}(\mathbf{x}_{1},\mathbf{x}_{2})
=
\sqrt{(\mathbf{x}_{1}-\mathbf{x}_{2})^{T}\cdot\bar{\mathbf{C}}^{-1}\cdot(\mathbf{x}_{1}-\mathbf{x}_{2})}
\end{flalign}
}
where $\bar{\mathbf{C}}$ denotes covariance matrix. As a result, PCC-KMC based on the Mahalanobis metric can produce a cluster of data points statistically proximal over the 5-dimensional color and pixel space. Given a set of L number of observations $(\mathbf{y}_1,\mathbf{y}_2,…,\mathbf{y}_L )$, each observation includes RGB color values at a $r^{th}$ and $c^{th}$ pixel, i.e., $\mathbf{y}_i=\left\{R_i,G_i,B_i,r_i,c_i\right\}$, where $L$ is equal to the total number of pixels of the possibly symptomatic regions which were labeled by 0 from the initial screening.
Furthermore, PCC-KMC partitions $L$ observations into a given number $K (\leq L)$ sets, denoted as $\mathbf{S}=\left\{S_1,S_1,…,S_K \right\}$ in such a way of minimizing the total sum of the Mahalanobis distances between observations and centroids of the clusters, such as
{\color{black}
\begin{flalign}
\underset{\mathbf{S}}{\arg\min}
\sum_{i=1}^{K}\sum_{\mathbf{y}_{j}\in S_{i}}
\sqrt{(\mathbf{y}_{1}-\mathbf{y}_{2})^{T}\cdot\bar{\mathbf{C}}^{-1}\cdot(\mathbf{y}_{1}-\mathbf{y}_{2})}
\end{flalign}
}
where $\boldsymbol{\mu}_i$ is the mean of points $S_i$.

To examine the performance of PCC-KMC in accurately extracting disease progression of ShB in Jasmine 85, segmentation patterns of PCC-KMC and KMC (k = 5) were compared (Fig. 3B). KMC generated clusters which contained pixels which were similar in terms of color (RGB values) regardless of the location of the pixels in the image (Fig. 3C). On the contrary, PCC-KMC clustered regions in terms of color similarity and spatial proximity, as reflected by each cluster where stem regions consisting of pixels of simliar colors which were also in close proximity were grouped together (Fig. 3D).

\subsubsection{Testing the performance of PCC-KMC to other existing methods}
The performance of PCC-KMC was further evaluated by comparing it to other segmentation approaches using a different ShB-infected sample image (301056): (i) k-means clustering over RGB color space with the Euclidean metric (KMC), (ii) k-means clustering over 5-dimensional color and pixel space with the Mahalanobis metric (PCC-KMC), (iii) mean-shift clustering \cite{ComaniciuMeer2002} over 5-dimensional color and pixel space with the Mahalanobis metric (PCC-MSC), and (iv) simple linear iterative clustering (SLIC) (Fig. 4A). The number of clusters was pre-set as 20 and bandwidth as 1 since the use of KMC, PCC-KMC and SLIC requires the number of clusters whereas PCC-MSC requires the bandwidth information or window size.
\begin{figure}
\centering
\includegraphics[width=\linewidth]{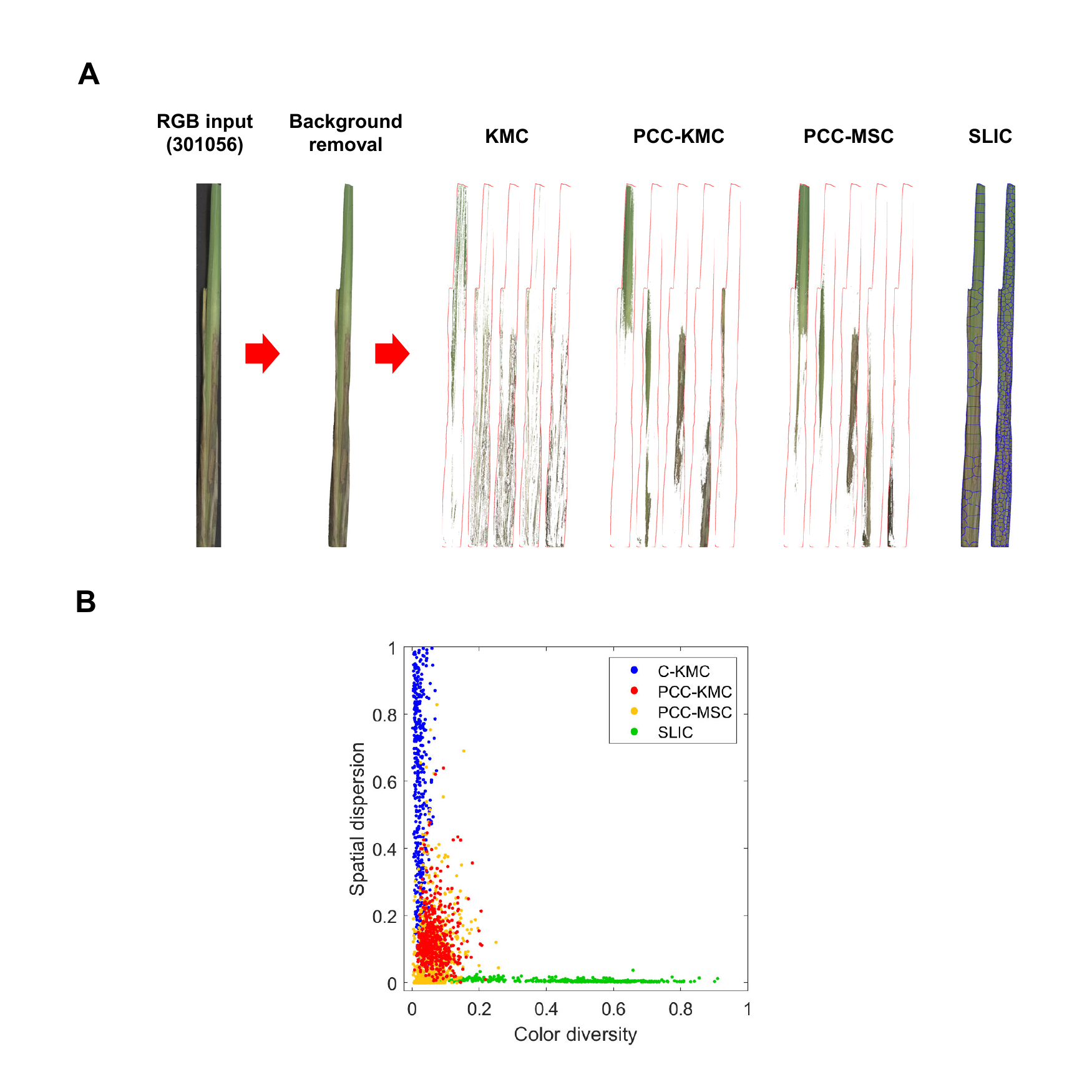}
\caption{Comparison of the performance of various image segmentation methods. {\bf A}, Illustration of representative clusters produced by KMC, PCC-KMC, PCC-MSC, and SLIC for a sample stem (301056) RGB image (k =20). {\bf B}, Clusters generated by the four different segmentation methods distributed based on color diversity and spatial dispersion axes for total 26 sample stem images.}
\label{Figure4}
\end{figure}
An overall comparison of the performance of each segmentation method was performed to illustrate the extent of color diversity versus spatial dispersion for all clusters obtained from segmenting 26 selected ShB-infected RGB images (Fig. 4B). Here, for each cluster, the color diversity is defined by
\begin{flalign}
\text{color diversity}
=
\text{tr}(\bar{\mathbf{D}}_{c} )/\text{tr}(\bar{\mathbf{D}}_{t})
\end{flalign}
where $\text{tr}(\cdot)$ denotes the trace that evaluates the sum of diagonal entries of a matrix and $\bar{\mathbf{D}}_{c}$ and $\bar{\mathbf{D}}_{t}$ are 3 by 3 diagonal matrices whose elements correspond to eigenvalues of covariance matrices of RGB color values of data points in a cluster and in all clusters, respectively.
The smaller the color diversity, the purer color quality a cluster has. Similarly, the spatial dispersion can be defined by
\begin{flalign}
\text{spatial dispersion}
=
\text{tr}(\bar{\mathbf{E}}_{c} )/\text{tr}(\bar{\mathbf{E}}_{t})
\end{flalign}
where $\bar{\mathbf{E}}_{c}$ and $\bar{\mathbf{E}}_{t}$ are 2 by 2 diagonal matrices whose elements are eigenvalues of covariance matrices of row and column pixel index of data points in a cluster and in all clusters, respectively. The larger the spatial dispersion becomes, the spatially less-local pixels comprise a cluster.

{\color{black}
\subsubsection{Convolutional neural network-based automatic ShB symptom classification}
Each of the clusters generated after being processed by PCC-KMC needs to be classified into ShB symptomatic or non-symptomatic cluster.
In order to automate this classification step, a convolutional neural network (CNN) was employed.
142 ShB-infected rice stem images were PCC-KMC processed wherein each image was segmented into 20 clusters. A total of 2840 clusters were classified into one of four classes: (i) ShB lesion, (ii) gradation (symptomatic region surrounding ShB lesion), (iii) green tissue (non-symptomatic region) and (iv) edge (border of plant samples, resulting from imperfect background removal).
RGB color histograms of all images were also included as input dataset to train the CNN classifier. However, due to inhomogeneity of images in terms of brightness, contrast, gamma, we normalized the RGB color histogram of each cluster in a range from 0 to 1 (double format). As a result, the resulting aggregate input dataset was stored in a 3-dimensional array sized by $1000\times3\times 2840$. Default functions provided by MATLAB \cite{MATLAB2019}, such as \texttt{trainNetwork} and \texttt{convolution2dLayer}, were utilized to train the CNN classifier and design a 2-dimensional convolutional layer.
For optimal performance of the CNN, we applied the 85:15 ratio rule between training and testing sets and manually classified clusters for the training set (i.e., the ground truth). The detailed parameters are listed in Table \ref{table:parameter}.
\begin{table}[ht]
\caption{Parameters to build a CNN architecture} 
\centering 
\begin{tabular}{c c} 
\toprule
filter kernel & uniform \\ 
filter size & $10\times 3$ \\
initial learning rate & 0.001 \\
training option & SGDM with 0.9 momentum \\
maximum epochs & 30 \\
minimum batch size & 100 \\
\bottomrule
\end{tabular}
\label{table:parameter} 
\end{table}
Our CNN classifier consists of a single 2-dimensional convolution layer, relu-layer, fully-connected-layer sized by 4, and softmax-layer.
The 2-dimensional convolution layer employs a uniform filter sized by $10\times 3$.
The CNN architecture was trained with the stochastic gradient descent with momentum (SGDM) with momentum 0.9, initial learning rate 0.001, maximum epochs 30 (An epoch is a full pass through the entire data set), and minimum batch size 100.

Fig. 5A and Fig. 5B present the training progress plot, depicting accuracy and loss versus epoch, respectively. The final accuracy and loss were recorded as 92.00\% and 0.2117, respectively.
\begin{figure}
\centering
\includegraphics[width=.8\linewidth]{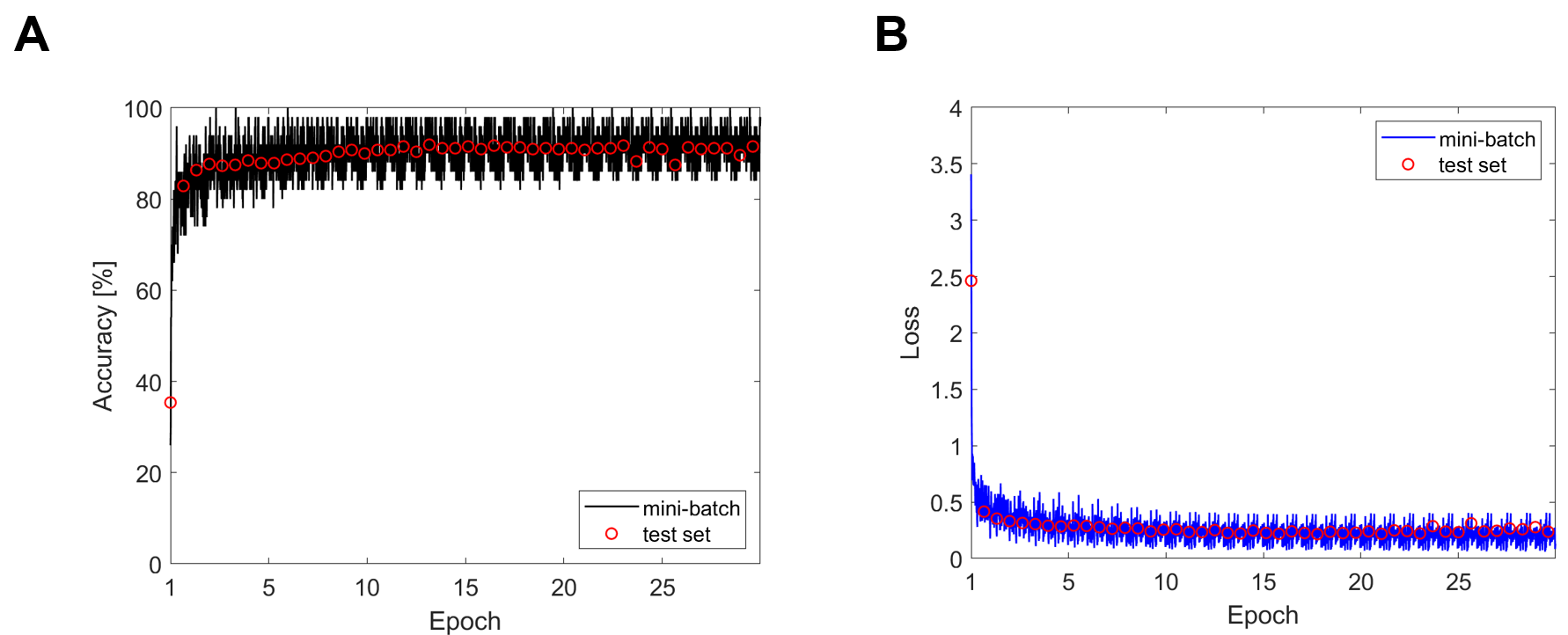}
\caption{Training progress plot of a convolutional neural network. {\bf A}, Mini-batch accuracy [\%] versus Epoch for training and test sets (black solid line and red circle, respectively). {\bf B}, Mini-batch loss versus Epoch for training and test sets (blue solid line and red circle, respectively).}
\label{Figure5}
\end{figure}

}

\subsubsection{Post-processing to quantify disease severity}
For each symptomatic binary mask after CNN classification, we performed post-processing to quantify the disease severity in terms of absolute and proportion lesion height and area. The absolute lesion area was calculated by scaling the total number of pixels at which a binary value is equal to unity. In our experiments, since the actual dimension and pixel size of a scanned image were $\left(21.77 cm \times 31.29 cm\right)$ and $\left(5135 \times 7383\right)$, respectively, the absolute area for each pixel was calculated as $(21.77×31.29)/(5135×7383)=1.7968×10^{-5}$ cm\ensuremath{^2} and the absolute length of one side of each pixel was calculated as $\sqrt{1.7968×10^{-5}}=4.2389×10^{-3}$ cm.
The total number of the pixels from a binary mask was counted using the MATLAB bwarea function. In addition, proportion diseased area was obtained by dividing the absolute symptomatic area with the absolute area of the stem. Likewise, absolute lesion height was calculated by scaling the number of pixels from the base of the stem to the disease lesion farthest from the base.
Since noises or incorrectly classified symptomatic regions were distributed over higher parts of stems, the measure of lesion height was significantly  altered even if the portion of the artifacts is extremely small compared with the total disease region. To avoid this situations, we calculated a cumulative distribution function (CDF) of symptomatic area with respect to stem height. Then, for a given height, CDF returned a value ranging from 0 to 1, representing the relative portion of symptomatic region. We visualized this CDF using a bar graph, which is opaque up to 0.9 of CDF and transparent from 0.9 to 1 of CDF with the extent of the CDF value in Fig. 6A. As such, we can compare the performance of image processing-based, disease quantifications to visual rating. As a note, the reference (visual rating) is the method which a human rater visually measured the lesion height on the original RGB image of the ShB-infected stem using MATLAB.
\begin{figure}
\centering
\includegraphics[width=\linewidth]{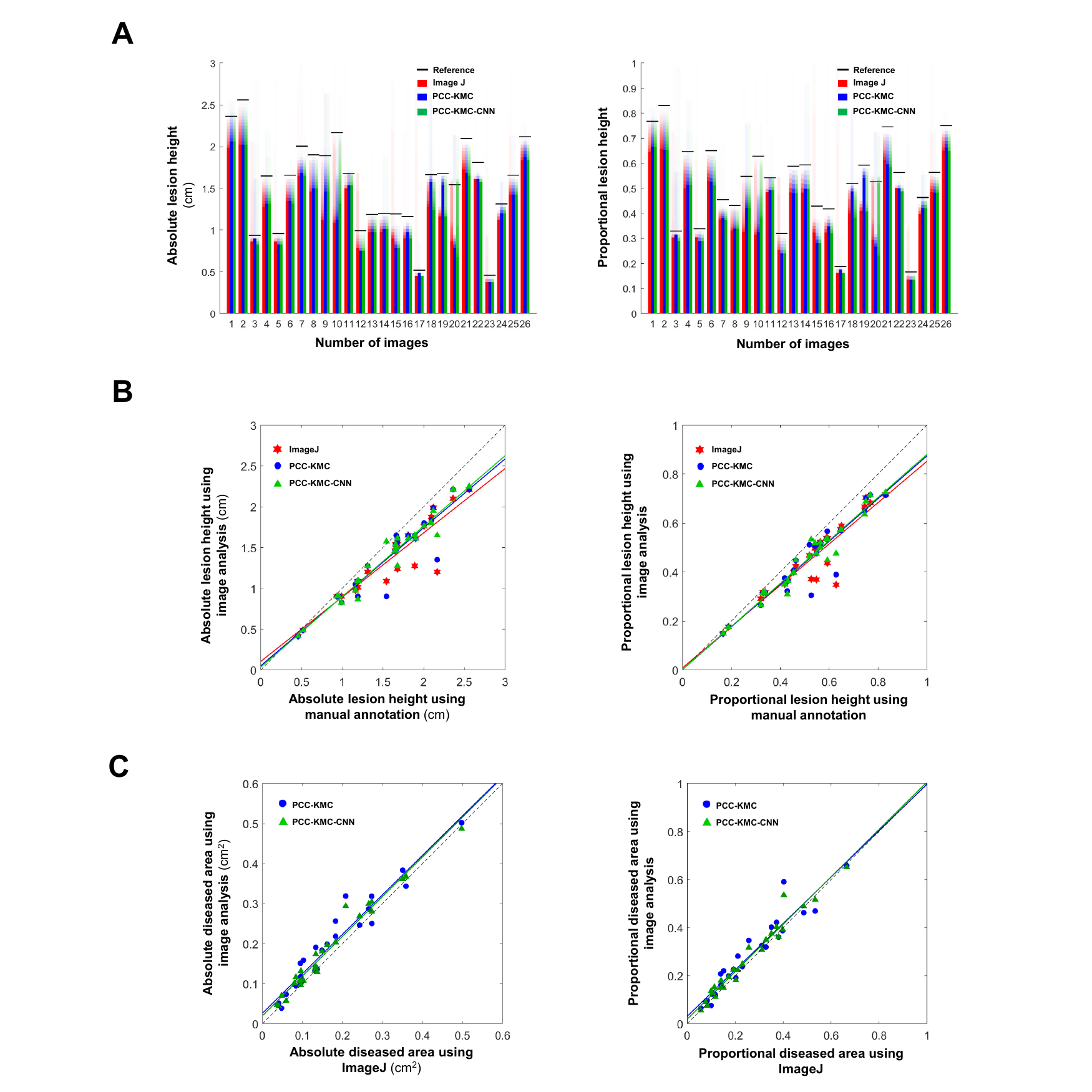}
\caption{Comparison of lesion height and diseased area measurements obtained from manual annotation, ImageJ, PCC-KMC and PCC-KMC-CNN. {\bf A}, Bar graph depicting absolute and proportional lesion heights of 26 ShB-infected rice culm images. Each bar is visualized by being transparent where the CDF value of the detected diseased region starts from 0.9, in which the extent of the transparency depends on the CDF value. {\bf B}, Agreement between lesion heights obtained using manual annotation and those obtained using ImageJ, PCC-KMC and PCC-KMC-CNN. {\bf C}, Agreement between diseased area obtained using ImageJ, PCC-KMC and PCC-KMC-CNN.}
\label{Figure6}
\end{figure}

\subsection{Evaluation of the accuracy of PCC-KMC and PCC-KMC-CNN}
To validate the results obtained by the PCC-KMC algorithm, we measured lesion height of ShB progress on original RGB images using MATLAB to serve as the reference of lesion height measurement. As for the reference measurement for symptomatic disease area, we utilized color thresholding feature in ImageJ \cite{Schneider2012}, wherein we obtained a symptomatic binary mask which was then post-processed in MATLAB to extract lesion height and area. ImageJ has been widely used as a gold standard in image-based, plant disease quantification \cite{Bock2020,Peressotti2011}, hence served as the actual value in comparing the symptomatic diseased area generated by PCC-KMC and PCC-KMC-CNN. Moreover, color thresholding was performed in LAB color space with ad-hoc thresholding values for every sample image.
Finally, using 26 sample images which were not a part of the training set in the CNN classifier, we compared results (lesion length and the area of diseased region) generated by visual assessment, ImageJ, PCC-KMC with manual labeling, and PCC-KMC-CNN classifier. The accuracy and precision of the measurements by PCC-KMC and PCC-KMC-CNN were tested using the Lin’s concordance correlation coefficient  \cite{Lin1989}.

\section{RESULTS}
\subsection{Performance of PCC-KMC}
Fig. 4A compares the segmentation results by KMC, PCC-KMC, PCC-MSC and SLIC.
KMC tends to produce clusters wherein pixels in each cluster possess higher color correlations and less spatial correlations.
In contrast, both PCC-KMC and PCC-MSC effectively separated symptomatic and non-symptomatic clusters.
In the case of SLIC, the resulting clustered regions were spatially over-localized while RGB color values were less correlated. Although the performance was better by increasing the number of clusters ($N_k=2000$ as seen in the right stem image under SLIC), it was not an efficient method feasible for training a convolutional neural network since one had to label too many clusters in each image.
PCC-KMC and PCC-MSC produced clusters which had the smaller spatial dispersion and color diversity which coincided with the result shown in Fig. 4A. Collectively, since ShB symptomatic regions were often spatially localized with distinct colors, the PCC-KMC and PCC-MSC segmentation methods were appropriate for our purpose to correctly extract features of symptomatic regions (Fig. 4B). Moreover, KMC generated clusters that only had the smaller color diversity but the larger spatial diversity and the opposite trend were observed for SLIC. Finally, KMC and SLIC were not suitable for accurate feature extraction nor quantification of plant disease since segmented regions did not represent ShB symptomatic regions.

\subsection{Accuracy of ShB severity measurements obtained from visual manual annotation, ImageJ and PCC-KMC and PCC-KMC-CNN}
For all sample images, we observed that reference results were lower and upper bounded by the CDF values 0.89 and 0.97, respectively (the average CDF value was around 0.93). This result suggests that visual rating and image processing-based disease estimates were within a close range of values (Figure 6A).
Agreement ($\rho_{c}$) of absolute lesion height (cm) among different methods to visual measurement was highest for PCC-KMC ($\rho_{c} = 0.87$), followed by ImageJ ($\rho_{c}= 0.83$) and PCC-KMC-CNN ($\rho_{c} = 0.78$) (Table 2, Fig. 6B).
\begin{table}
\centering
  \caption{Summary of the performance of PCC-KMC and PCC-KMC-CNN in 26 ShB-infected rice culm RGB images compared to visual manual annotation and ImageJ measurements.}
  \label{Table2}
  \includegraphics[width=\linewidth]{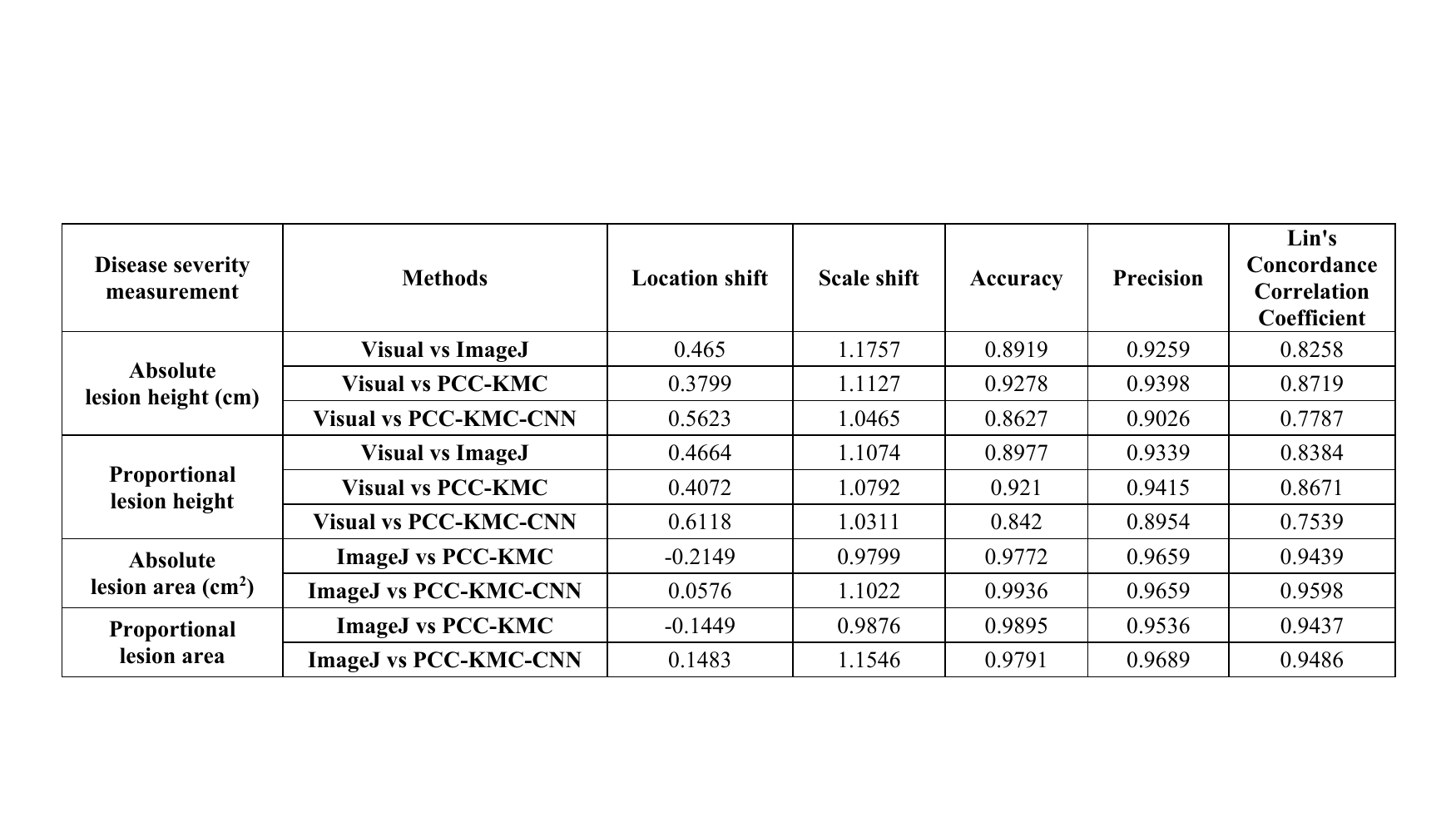}
\end{table}
The same trend was observed for the agreement of proportional lesion height among the different measurement methods. For diseased lesion area (cm2), PCC-KMC-CNN was slightly more accurate ($\rho_{c}=0.96$) than PCC-KMC ($\rho_{c}=0.94$) upon comparing their measurement values to that of ImageJ. Similar pattern was also observed for proportional diseased area (Fig. 6C). Moreover, ImageJ, PCC-KMC and PCC-KMC-CNN showed a good correlation between absolute lesion height and absolute diseased area (r = 0.80 to 0.86) (Table 3, Fig. 7).
\begin{table}
\centering
  \caption{Correlation analysis of ShB lesion height and disease area of 26 ShB-infected rice culm RGB images using ImageJ, PCC-KMC and PCC-KMC-CNN.}
  \label{Table3}
  \includegraphics[width=\linewidth]{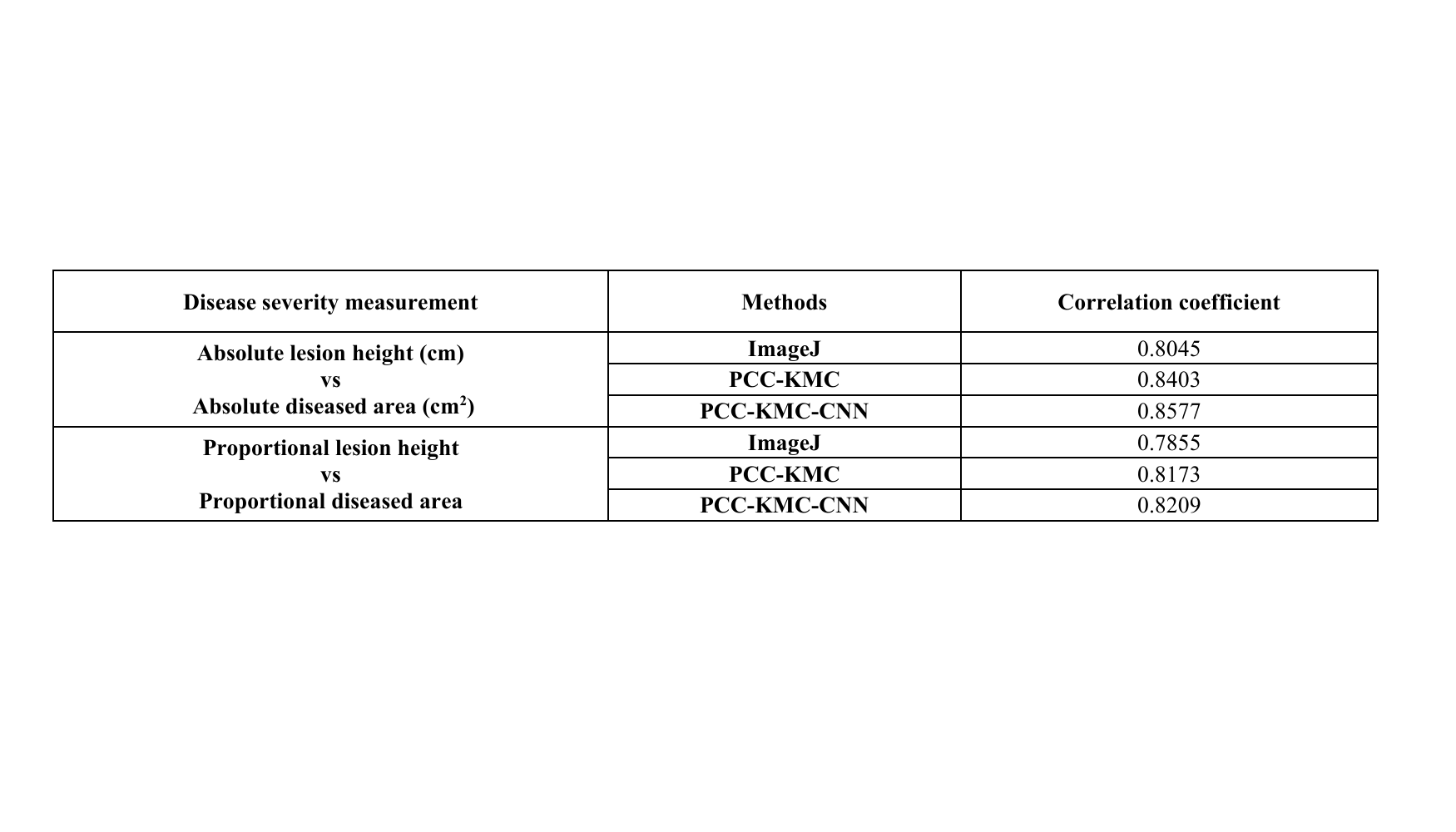}
\end{table}
\begin{figure}
\centering
\includegraphics[width=\linewidth]{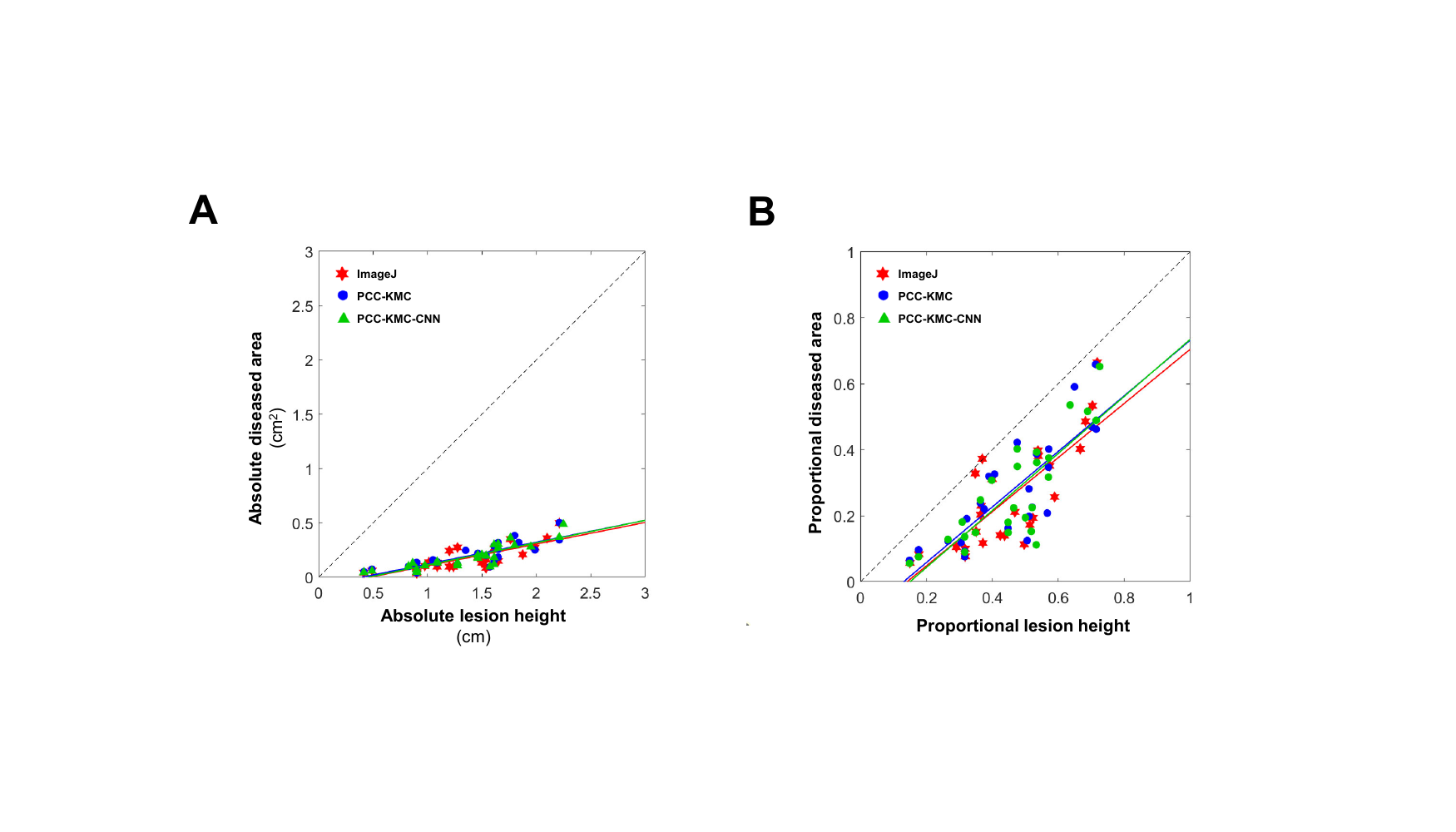}
\caption{Correlation of ShB lesion height and symptomatic disease area. Relationship of  lesion height (cm) and diseased area (cm2) in terms of {\bf A}, absolute and {\bf B}, proportional values. }
\label{Figure7}
\end{figure}

\section{Discussion}.
Moreover, we introduced a convolutional neural network (CNN) for automatic classification on clusters produced by the PCC-KMC so that the present digital image quantification method becomes fully automatic.
The performance and feasibility of the unsupervised (PCC-KMC) and supervised (convolutional neural network or PCC-KMC-CNN) learning were also evaluated to test their potential to be utilized for efficient extraction of disease phenotype information from RGB images.
A high correlation between PCC-KMC or PCC-KMC-CNN and the visual-based, manual annotation method was obtained.
Our study has provided a new strategy for disease severity assessment of ShB and other plant diseases.

Furthermore, instead of standard Euclidean distance, we measured the spatial closeness of pixels by the Mahalanobis metric so that it can remain unitless and scale-invariant, taking into account the correlations of the data set.
Hence, the present PCC-KMC algorithm is more suited to segment ShB symptomatic regions on RGB images of infected rice culms and quantify ShB severity (lesion length and diseased area).

A wide variety of methods are being used to describe the ShB resistance of germplasms, including disease indices \citep{Jia2007,Shah2008}, 0 to 9 scale \citep{Prasad2008} and combination of lesion length of three adjacent leaves \citep{Prasad2008}, which poses challenges for the comparison of the results from different studies. In the field, pathogen inoculation methods such as liquid inoculum injection \cite{Sato2004}, mycelial toothpick inoculation \cite{Zou2000} and solid inoculum injection are widely utilized, whereas for under controlled conditions, microchamber assay \cite{Jia2007}, mist-chamber \cite{Jia2013} and detached leaf method \cite{Venu2007, Jia2013} are used. However, previous studies reported that different evaluation methods may generate different ShB disease phenotypes of the same genotypes \cite{Hossain2014}. For instance, two moderately resistant cultivars such as Tetep and Teqing, were found to be resistant in mist-chamber inoculations, however, under micro-chamber conditions, enhanced disease scores were observed in the two cultivars \cite{Hossain2014}. Furthermore, the environment where the experiment is performed \cite{Hossain2014}, morphological characteristics of rice \cite{Groth1992,  Hossain2016} as well as the age of the plants being tested \cite{Hossain2014,  Jia2007} are factors influencing the resistance of rice cultivars. For instance, susceptible rice cultivar, Cypress, surprisingly showed resistance once inoculated in field conditions \cite{Jia2007}.

For polygenic disease resistance screenings such as for ShB, a wide {\color{black}range} of susceptibility profiles among different rice cultivars can be observed \cite{Groth2008,  MarchettiBollich1991,  ShaZhu1990, Jia2007, Hossain2014}. Moreover, the characteristic necrotic, water-soaked, irregular-margined ShB disease lesions occurring on leaf sheaths and blades of rice plants \cite{BannizaHolderness2001,  Ogoshi1987,  Yellareddygari2014} make it challenging to determine the exact lesion height in infected plants.

However, another devastating symptom caused by ShB is stem lodging, which occurs at the lower internodes of rice plants \cite{Wu2012, HoshikawaWang1990} and consequently causes weakening of the rice stems. This leads to the interruption of transportation of water and nutrients through the xylem and phloem \cite{Kashiwagi2005} and negatively affects the photosynthetic ability of the plant, leading to poor grain filling and biomass production of rice plants \cite{Lang2012,  Setter1997, RushLee1983}. Basu and colleagues \cite{Basu2016} reported the significant difference in lesion development between rice varieties Swarna (susceptible) and Swarnadhaan (tolerant) in terms of lesion length with the increase in time (days post inoculation). In addition, microscopic examination of tolerant and susceptible ShB-infected host tissue, wherein greater hyphal density, number of infection cushions and microsclerotia were observed in the susceptible variety, suggesting the preferential colonization of R. solani affecting the differential outcome of ShB disease symptoms in these varieties \cite{Basu2016}. Since ShB symptoms tend to first appear near the stem base and then extend towards the rice canopy \cite{HashibaKobayashi1996,  Ogoshi1987}. The tissue damage and obstruction of water and nutrient transportation to the upper regions of the plant affects rice production \cite{HoshikawaWang1990}. Moreover, ShB infestation was reported to reduce stem breaking resistance leading to stem lodging \cite{Wu2012}. Hence, these limitations emphasize the need to develop less subjective approaches capable of providing high-throughput yet accurate measurements of the quantitative ShB resistance phenotypes reflected by different rice germplasm. In addition, ShB severity and its consequent resistance phenotypes must not only be measured in terms of lesion progression but also in terms of diseased area in order to have a complete assessment of ShB resistance, which can be overcome with the use of image-based disease detection and quantification as shown in this study.
In a digital image, except for boundaries, the closer pixels generally possess the higher correlation such as the pixels in the immediate neighborhood will share the similar features. In addition to color information, spatial relationships of neighboring pixels can be also of great aid in imaging segmentation, particularly, for plant diseases exhibiting gradual dispersion of diseased area or necrosis such as ShB. Moreover, clustering considering the spatial information can yield more homogeneous clusters less noises and reduce the spurious blobs \cite{Chuang2006}. Numerous studies employing k-means clustering, means-shift clustering and SLIC showed enhanced performance of image segmentation by including the spatial information in a feature space \cite{Achanta2012, ComaniciuMeer2002, Luo2003, Chuang2006}. Basically, k-means and SLIC require the non-physical parameter, the number of clusters, whereas mean-shift operates with the physical parameter, window size or bandwidth; hence, mean-shift is often preferred for image segmentation of general purposes. However, the computational complexity of these methods is linearly proportional to the number of iterations while mean-shift is relatively expensive due to the quadratic order of the number of iterations. Based on our results, PCC-KMC and PCC-MSC seem to provide similar results. In fact, mean-shift clustering is a general and application-independent tool without setting the number of clusters. However, its limitation is that it requires the selection of an optimal bandwidth, which is a non-trivial task and often computationally expensive than k-means clustering and PCC-KMC. Furthermore, SLIC divides the search region into a set of small areas, which are relatively regularly-shaped, it may not be efficient and accurate to segment arbitrarily-shaped and -scaled symptomatic regions. Hence, in this paper, we decided to use k-means clustering with the spatial information to facilitate efficient processing of high number of sample images to extract data of statistical and biological significance and with the ease of implementation.
Acquisition of rice culm images while the plant is still planted in the pots or in the field can reduce the accuracy of the image-based disease measurements due to numerous factors such as variation angle of image acquisition, illumination, distance between the sensor and plant target. Hence, in order to acquire the sample images in a systematic and uniform manner, the use destructive sampling was inevitable in this study. However, some limitations to this approach may exist since rice culms are not flat and thin as paper and shadows can be casted over the margins of the plant upon scanning of the stems. In addition, to prevent the sample to dry and become brittle, it is recommended that the samples should be scanned right after cutting out from the plant or be place in a Ziploc bag with a damp paper towel in it.
Lin’s concordance correlation coefficient has been used to assess the accuracy of plant disease quantification methods by comparing the estimated and actual values of disease severity measurement \cite{Lin1989, StewartMcDonald2014, Bock2008}. We observed high levels of accuracy and precision of lesion height upon comparing PCC-KMC to visual measurements and for PCC-KMC and PCC-KMC-CNN to Image J, which supports that the PCC-KMC method mimics the human visual measurements and the gold standard, respectively. This suggests that PCC-KMC is effective in detecting ShB in RGB images of ShB-infected rice culms and automation using this method of image segmentation is possible. With the existing ImageJ software which requires manual in thresholding for every image being analyzed, the automated PCC-KMC-CNN system image will serve as a foundation for constructing a pipeline of high-throughput ShB resistance screening experiments. As for future work, we recommend the testing of PCC-KMC-CNN using a larger number of sample images with pre-determined visual-based resistance phenotype data (resistant or susceptible), which can be incorporated in the CNN and provide qualitative resistance phenotypes automatically for each image.

In conclusion, we developed an image-based, disease quantification system capable of accurate detection of ShB resistance reactions at early disease development. The performance of both PCC-KMC and PCC-KMC-CNN was confirmed with visual-based estimations of ShB disease symptoms. This method provides reliable and precise phenotypic data of ShB disease for resistant germplasm screens and breeding programs. Furthermore, the method could be easily adapted for evaluating similar diseases in other plants.

\section*{Acknowledgements}
This work was supported by a fellowship awarded to D.-Y. Lee by the Monsanto Beachell-Borlaug International Scholarship Program (MBBISP).

\section*{Authors' contribution}
DL, DN, GL conceived and designed the experiments, DL and DN conducted the experiments, developed the algorithm and wrote the manuscript, YH and GL reviewed and edited the manuscript.

\bibliographystyle{vancouver}
\bibliography{PCC_KMC_biography_v1}


\end{document}